# Quantifying Flow State Dynamics: A Prefrontal Cortex EEG-Based Model Validation Study.

## Unveiling the Prefrontal Cortex's Role in Flow State Experience: An Empirical EEG Analysis.

Gianluca Rosso[1], Raffaella Ricci[2], Lorenzo Pia[2], Giovanni Rebaudo[3], Michele Guindani[4], Alberto Marocchino[5], Giorgio De Pieri[6], Andrea Filippo Rosso[5]

[1]Department of Economics & Statistics "Cognetti de Martiis", University of Turin, Turin, Italy; [2]Department of Psychology, University of Turin, Turin, Italy; [3]Department ESOMAS, University of Turin, Turin, Italy; [4]Department of Biostatistics, Jonathan and Karin Fielding School of Public Health, University of California, Los Angeles (UCLA), Los Angeles, CA 90095, USA; [5]Sporthype, Turin, Italy; [6]Technical Director of Golfus Performance Center, Head Coach Saudi Arabia National team Men's.

## INFO

This work is a preprint and has not been certified by peer review, 2025
Version posted: 06, 30, 2025

Conflict of Interest Statement:
This work was supported by Sporthype. Some of the authors are employees of Sporthype, which contributed technical expertise essential to the completion of this study. The authors declare that Sporthype did not influence the study design, data analysis, or interpretation of the results beyond their technical role.

Scientific Editor: Isabella Pecetto

Correspondence to:
glr@gianlucarosso.com

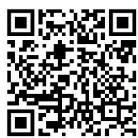

## ABSTRACT

This article aims to explore the optimization of mental performance through the analysis of metrics associated with the psychological state known as flow. Several clinical studies have shown a correlation between the mental state of flow (characterized by deep and relaxed concentration and high psychophysical efficiency) and brain activity measured through electroencephalography (EEG). This study confirms such a correlation, focusing in particular on the sports field, where the flow state tends to occur more frequently. Among athletes, this state contributes to more effective stress management and greater clarity in defining goals. To conduct the study, Sporthype developed proprietary software that integrates several predictive models, in particular the Flow State Index (FSI), implemented within the Holytics system. An analytical protocol was established, including mental exercises and data collection sessions using the portable EEG device Muse, accompanied by a questionnaire to gather athletes' subjective perceptions of their mental state. The experimentation was carried out with certified professional golfers, thus allowing the empirical validation of the models in real scenarios. The analysis involved not only technical movements, but also fundamental cognitive components such as mental imagery and concentration, with the aim of building a preparation routine that enhances the athlete's awareness and ability to access the flow state. The results revealed a significant alignment between the EEG data and the subjective experiences reported in the questionnaires, confirming the feasibility of detecting the flow state through prefrontal cortex activity. Furthermore, the psychological exercises included in the study protocol showed a tangible positive effect in enhancing flow during athletic performance. From a neurophysiological point of view, the flow state is associated with reduced activity in the dorsolateral prefrontal cortex—a phenomenon known as transient hypofrontality. This condition reduces negative internal dialogue, thus improving cognitive and motor efficiency. Proper synchronization between the left and right hemispheres also promotes the integration of logic and intuition, both of which are essential in disciplines such as golf. In summary, flow improves performance through a more harmonious synchronization between mind and body. Although golf was the main context of the experimentation, the mathematical models developed within Holytics were designed to be applicable to a wide range of sports. In addition to golf, preliminary tests have been conducted in other sports such as tennis, as well as in non-sport contexts, including gaming and mental training practices such as mindfulness, concentration, and visualization.

**INDEX.**

### 1. Introduction.

One of the most recognized concepts in human performance studies is the psychological state known as flow. This mental condition is characterized by deep engagement in the task and a marked reduction in self-referential thoughts, such as rumination and excessive self-consciousness. Flow is frequently observed in high-performing individuals, like athletes, musicians, or researchers, who are fully absorbed in their tasks to reach optimal levels of achievement. However, such states are not limited to extraordinary contexts; they can also arise during everyday activities, whether at work or during leisure. For instance, a video game player might spend several uninterrupted hours engaged in play, feeling neither tired, hungry, nor bored. Entering a flow state is often linked to a heightened sense of purpose, personal fulfillment, and emotional uplift. As a result, it contributes meaningfully to an individual's overall well-being.

The flow state [Csikszentmihalyi, 1990, 2014] represents a highly complex aspect of the human mind, playing a crucial role in executing various activities. The flow state can be assessed using an electroencephalography (EEG), provided the device is worn in accordance with standardized protocols: EEG can be a valuable tool for assessing flow when the device is used under well-defined protocols that include proper electrode placement, artifact management, environmental control, and rigorous data analysis. These measures help ensure that the brain activity recorded truly reflects the neural dynamics of the flow state.

Sporthype has developed a system for analyzing the flow state that integrates the use of an EEG device, focusing on prefrontal cortex activity, performing data analysis, and providing real-time feedback. To address potential concerns regarding the reliability of the EEG methodology, this article will first provide a detailed description of the technology and the specific device used in the study. Subsequently, the analytical model applied to process and interpret the recorded EEG data will be presented.

An evaluation of the model is then provided using an empirical system that compares the results with the flow sensation experienced by the subjects involved in the test. Finally, practical guidelines are offered for managing the tests, allowing users to engage in mental training related to the flow state and track their progress over time through software distributed via Holytics system. In addition, a recommendation system will be described that offers practical advice to enhance the flow state and achieve an optimal level of performance.

### 2. The Flow State: A Neurophysiological Perspective on Deep Task Engagement.

Flow state, often described as being "in the zone," is a mental state of complete absorption and focus in an activity. It was first conceptualized by psychologist Mihály Csíkszentmihályi [Csíkszentmihályi, M., 1975,

Beyond Boredom and Anxiety: Experiencing Flow in Work and Play]. Below is a detailed explanation of flow state, including its characteristics, mechanisms, and benefits.

Flow is a psychological state of optimal experience in which an individual is fully immersed in an activity, experiencing energized focus, deep engagement, and intrinsic enjoyment. It is characterized by a seamless merging of action and awareness, where time seems to be distorted, either decelerating or accelerating, and self-consciousness fades.

Csíkszentmihályi identified nine components that define the flow experience:

1. Challenge-Skill Balance: tasks must be challenging but achievable, ensuring engagement without overwhelming or boring the individual.
2. Action-Awareness Merging: complete absorption in the activity eliminates distractions and self-conscious thoughts.
3. Clear Goals: a defined purpose provides structure and direction.
4. Immediate Feedback: real-time responses to actions help adjust performance and maintain focus.
5. Concentration on the Task: intense focus prevents attention from wandering.
6. Sense of Control: individuals feel in command of their actions and outcomes.
7. Loss of Self-Consciousness: concerns about oneself or external judgment disappear.
8. Transformation of Time: time perception changes—hours may feel like minutes or vice versa.
9. Autotelic Experience: the activity is intrinsically rewarding, pursued for its own sake rather than external rewards.

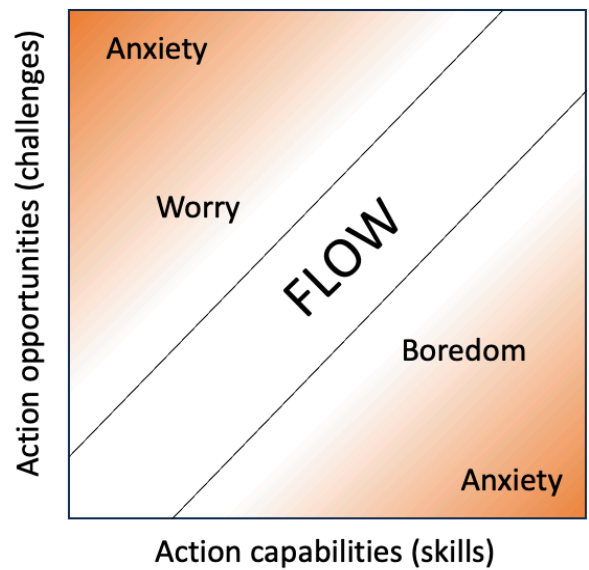

*Fig. 1 - Adaptated picture from the original model of the flow state (Csikszentmihalyi, 1975, p. 49)*

Three primary conditions must be met to enter flow:

I. Clear Goals: the activity must have a clear structure and objectives.
II. Immediate Feedback: feedback helps adjust efforts to stay aligned with goals.
III. Balance Between Challenge and Skill: the task must match the individual's abilities, too easy leads to boredom, while too difficult causes anxiety. The "Experience Fluctuation Model" developed by Csíkszentmihályi (1997) illustrates this balance as central to achieving flow.

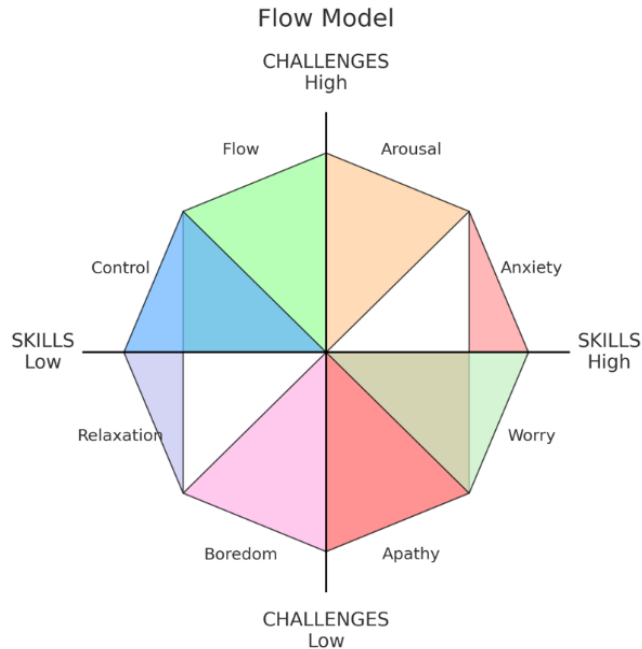

*Fig. 2 - Adaptated picture from Refined Flow Model (Csikszentmihalyi, 1997; Csikszentmihalyi, 2014, p. 201)*

Flow involves specific brain activity and neurotransmitters:

- Transient Hypofrontality Theory: during flow, parts of the prefrontal cortex responsible for self-reflection and critical thinking show reduced activity, allowing automatic processes to dominate.
- Neurotransmitters: dopamine (reward motivation) and norepinephrine (focus and energy) are crucial in sustaining flow.

Brain imaging studies have shown increased activation in areas supporting task demands while suppressing self-referential thought.

Flow offers numerous psychological and physiological advantages:

- Enhanced Performance: it boosts productivity in creative, athletic, and professional domains.
- Improved Well-Being: flow fosters happiness by creating intrinsically rewarding experiences.
- Stress Reduction: it serves as a coping mechanism for anxiety when engaging in enjoyable activities.
- Skill Development: sustained focus promotes mastery and learning.

While generally positive, flow has potential drawbacks:

- Overindulgence in certain activities (e.g., video games) can lead to neglect of other responsibilities.
- Hyperfocus associated with flow may sometimes hinder broader problem-solving or multitasking abilities.

To cultivate a flow state, the following conditions should be met:

- Engage in activities that align with existing skill levels while providing a slight challenge beyond the current comfort zone.

- Set clear goals and ensure immediate feedback during the task.
- Minimize distractions by creating an environment conducive to deep focus.
- Practice mindfulness or meditation to enhance concentration.

As mentioned above, flow can be experienced through various domains—from sports and creative pursuits to everyday tasks like cooking or gardening, making it accessible to everyone under the right conditions.

According to various EEG studies [Katahira K. Et al., 2018; K. K. Yoshida et al., 2018], the flow state is associated with specific brain activation patterns. In particular, Katahira K. et al. (2018) highlight how, during a mental arithmetic task, flow manifests as an increase in theta waves in the frontal region and moderate alpha activity in the frontocentral areas. This balance between activation and relaxation suggests an optimization of cognitive resources, facilitating concentration and optimal performance [Bauer C. et al. 2020].

### 3. Immediate Feedback: The Catalyst for Flow.

Immediate feedback (through a Flow State Index FSI intended as a KPI) is vital in achieving flow states, as it provides real-time information about performance, allowing individuals to adjust their actions and stay deeply engaged in the task at hand. This type of feedback bridges the gap between what people are doing (input) and the outcome (output), ensuring they can make real-time adjustments. It keeps individuals focused on the present moment, essential for entering a flow state.

Immediate feedback enhances learning and performance by enabling the prompt correction of errors and the reinforcement of effective behaviors. This facilitates more rapid skill acquisition and improved outcomes. It also supports sustained attention by minimizing distractions and reducing the self-doubt that may arise from delayed feedback. By offering clear and timely information on performance, immediate feedback lowers cognitive load and uncertainty, thereby allowing greater mental capacity for more complex tasks. Furthermore, it fosters motivation by affirming competence when outcomes are positive and by providing actionable guidance for improvement when they are not. This ongoing reinforcement contributes to sustained engagement and enthusiasm, which are essential for maintaining a state of flow.

To integrate immediate feedback into activities, it is advisable to develop internal feedback mechanisms[1]. This entails establishing clear benchmarks or performance standards and continuously monitoring progress in relation to these predefined objectives.

Creating structured feedback loops is another strategy. Use software tools that provide instant analytics, such as trackers for athletes. Set up regular check-ins with supervisors or teammates during projects to ensure that feedback is integrated into your workflow seamlessly.

Effective immediate feedback should be specific and actionable, addressing precise actions and offering straightforward suggestions for improvement. It should be timely and provided immediately after an action to maintain relevance and context. Balancing frequency is also important; too much feedback can overwhelm, while too little can leave gaps in understanding. The goal is to find the minimum amount needed to guide performance effectively without disrupting the workflow.

By incorporating immediate feedback into your activities, you can enhance your ability to focus, improve your skills more efficiently, and increase your chances of entering a flow state. This real-time guidance not only boosts performance but also fosters a deeper engagement with the task at hand, leading to a more fulfilling experience.

---

[1] Internal feedback mechanisms are self-regulatory processes that allow individuals to monitor, evaluate, and adjust their performance during an activity, without relying on external input (like a coach or supervisor). These mechanisms help a person stay aligned with goals and improve effectiveness in real time. Key Elements of Internal Feedback Mechanisms: Clear Standards or Goals, Self-Monitoring, Self-Assessment, Adjustment or Correction. These mechanisms are essential for building autonomy, supporting flow, and sustaining high performance, especially in tasks that are complex, creative, or done in solitude.

The flow state can be assessed using an EEG device, which must be worn according to specific guidelines.

## 4. EEG device: characteristics, scientific uses, and formal validations.

The need to easily analyze brain activity without using laboratory medical diagnostic systems has led several technology companies to develop portable EEG devices that differ in their functionalities and costs.

Depending on the intended use, the choice of EEG devices ranges from simpler models with fewer sensors to more advanced ones with extended capabilities.

In the context of mental performance analysis, when integrated into a more global psychophysical assessment of the individual, devices that analyze only the brain prefrontal cortex activity are generally considered sufficient. Indeed, shifting the analysis from a purely medical domain to one typically related to meditation requires stepping away from strictly diagnostic applications. Nevertheless, the data obtained is adequate for the purpose and is also of good quality [Krigolson and colleagues, 2017; Herman K. et al., 2021]. Furthermore, the benefits become evident when the data is processed in a global big data context along with other variables such as lifestyle, daily habits, diet, and nutraceuticals. This is especially true considering that most portable EEG devices are genuinely small and lightweight, and, above all, have a highly competitive and user-friendly cost.

This device is certified. Muse complies with safety and electromagnetic compatibility standards required for personal electronic devices. It has obtained CE certification (necessary for sale in the European Union) and FCC certification (Federal Communications Commission for the U.S. market). These certifications do not address scientific accuracy but ensure the device is safe for general use.

Scientific validation: Independent studies have demonstrated that Muse's EEG data, while less precise than high-density clinical systems, is sufficiently reliable for research and biofeedback applications. For instance, Krigolson et al. (2017) validated its capacity to detect neural signals such as alpha and theta rhythms in controlled ERP paradigms.

Muse has been used in numerous academic studies published in peer-reviewed scientific journals (for example Badcock N.A. et al., 2013; McCrimmon C., et al., 2017), attesting to the device's reliability for non-clinical applications. These studies often focus on:

- Stress and attention monitoring
- Mindfulness and meditation research
- Sleep and emotional analysis

Thus, the Muse device by Interaxon is primarily marketed as a tool for meditation and biofeedback but is also used in scientific research applications.

The device was conceptualized in 2003 and developed over the years, culminating in its commercialization in 2014. The Muse 2 device, launched in 2018, adds features such as heart rate measurement and includes an accelerometer for assessing body movement, particularly head movement. It has been shown that Muse can be used for ERP research[2], with the advantage of being economical and quick to set up. Moreover, it is widely used for a variety of other applications, ranging from health and wellness to scientific and medical research.

---

[2] ERP research, or Event-Related Potentials studies, are neuroscientific investigations that analyze the brain's electrical activity in response to specific stimuli, recorded through electroencephalography (EEG). These signals allow for precise observation of cognitive processes such as attention, perception,

Muse is worn over the ears and connects to a companion mobile app via Bluetooth.

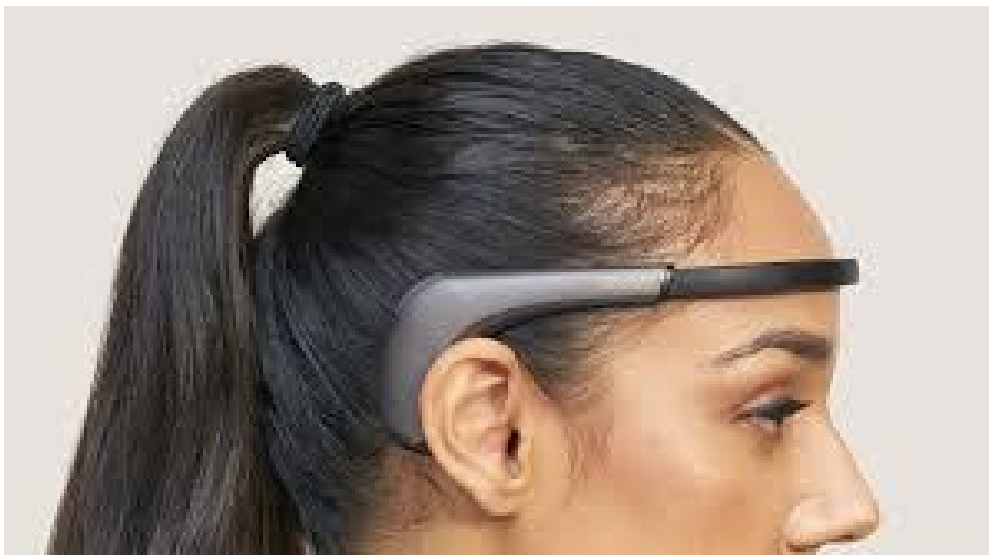

*Fig. 3 - Picture from the Interaxon web site.*

The use of Muse enables biofeedback functionality. The EEG sensors in the Muse 2 are dry electrodes that measure brain activity through brainwaves (alpha, beta, gamma, delta, theta).

Placement:

- Frontal area (FP1 and FP2) to monitor activity in the brain's frontal regions.
- Retroauricular area (TP9 and TP10), positioned near the ears.

Additionally, the Muse 2 includes:

- PPG sensor (Photoplethysmography): Measures heart rate by detecting blood flow variations in tissues, enabling real-time heart rate monitoring and supporting synchronization between breathing and heart activity, known as cardiac coherence.
- Accelerometer: Measures head movements along three axes (x, y, z).
- Gyroscope: Records changes in head orientation and rotation.
- Respiration sensor: Detects chest movements and records respiratory rate based on body micro-movements.
- Tilt and position sensors: Track head position and the degree of inclination.

---

memory, and language, indicating not only when but also how the brain reacts to certain events. ERPs are obtained by repeatedly presenting stimuli to participants and analyzing the brain's electrical response to each event. This research is highly valuable for understanding cognitive functioning in both healthy individuals and those with neurological or psychological disorders. A well-known example is the P300 component, which is associated with attention and decision-making processes.

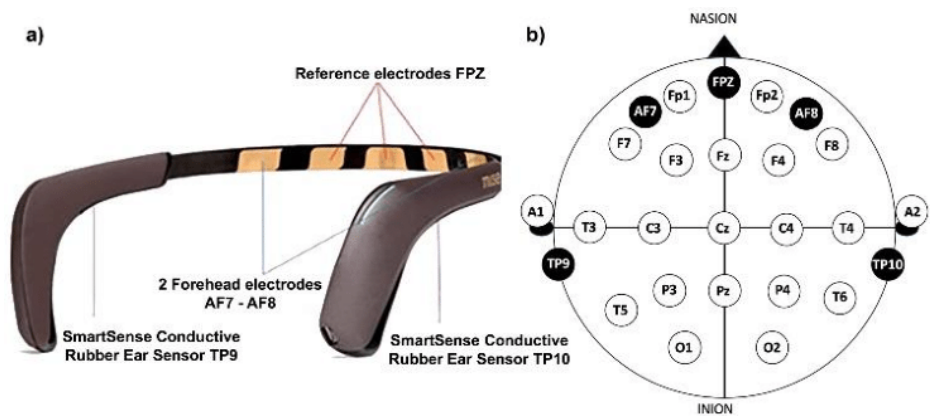

*Fig. 4 – Muse sensors.*

Compared to the first Muse, Muse 2 offers a broader range of sensors that not only monitor brain activity but also track physical and physiological parameters, making it ideal for a holistic view of mental and physical states during meditation or research.

The Muse S is an advanced version of the Muse device line, designed to integrate sleep monitoring with the meditation and biofeedback functionalities already present in earlier models. It is ideal for those seeking support throughout the day to relax and enhance meditation, as well as for obtaining data on sleep and nighttime rest.

Compared to the Muse 2, the Muse S features a softer, more flexible design made with fabric materials to ensure comfort during prolonged use, especially at night.

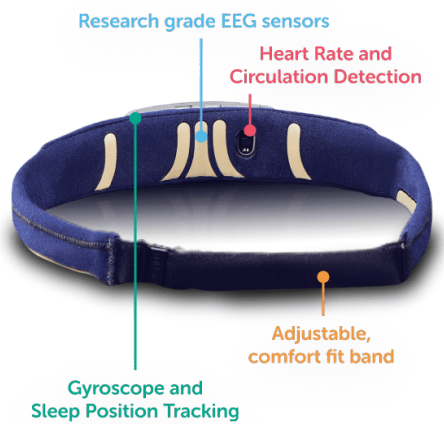

*Fig. 5 - Picture from the Interaxon web site.*

SPORTHYPE considers Muse 2 as standard equipment for Holytics and has conducted comparative tests with the Muse S, finding almost complete compatibility. The outputs are consistent between the two devices, allowing the HOLYTICS platform to manage either device interchangeably. The connection is made via Bluetooth, and over months of use, some differences in transmission systems have been observed. For this reason, preference when using the HOLYTICS platform is given to a mobile device, such as a smartphone. This ensures the complete system, comprising the Muse, smartphone, and HOLYTICS, is entirely portable.

There is no substantial difference between Android and iOS operating systems, although the latter requires a specific browser, which can be easily downloaded from the Apple Store.

The electrical signals from the prefrontal cortex are sufficient to determine stress, anxiety, cognitive phases, and focus because this brain region is deeply involved in regulating higher cognitive functions, emotions, and stress responses (Girotti M. et al. 2018; Negrón-Oyarzo I. et al 2016).

Below is a detailed explanation:

1. Role of the Prefrontal Cortex (PFC)

The PFC is responsible for complex cognitive functions, including:

- Emotional regulation: The PFC is closely connected to the limbic system, particularly the amygdala, which is involved in managing emotions like stress and anxiety.
- Attention and focus control: The dorsolateral PFC is critical for maintaining attention, concentration, and problem-solving.
- Decision-making processes: These are involved in planning, decision-making, and behavioral regulation.
- Stress self-regulation: The PFC helps modulate cortisol (the stress hormone) responses and inhibits impulsive or excessive reactions.

2. Electrical Signals and Neuronal Activity

Electrical signals (detected via techniques such as electroencephalography or EEG) reflect the synaptic activity of neurons. In the PFC, specific brainwave patterns are associated with different mental states (Priyanka A. et al 2016):

- Gamma Waves (30–100 Hz): Gamma oscillations represent the highest-frequency brain activity and are implicated in advanced cognitive functions, including learning, memory consolidation, and complex information processing. Elevated gamma activity is associated with heightened arousal, anxiety, and stress, while diminished gamma activity has been linked to neurodevelopmental and psychiatric disorders such as Attention Deficit Hyperactivity Disorder (ADHD)[3], depression, and learning impairments. Under optimal neurophysiological conditions, gamma waves facilitate sustained attention, sensory integration (e.g., olfactory, visual, and auditory modalities), conscious awareness, and higher-order perceptual processing.
- Beta Waves (12–30 Hz): Characterized by high frequency and low amplitude, beta rhythms are typically observed during wakeful, alert states and are associated with active cognitive engagement and logical reasoning. An optimal level of beta activity enhances concentration and cognitive performance. However, excessive beta activity may induce psychological tension, anxiety, hyperarousal[4], and difficulty relaxing, while reduced beta activity is correlated with symptoms of ADHD, diminished executive function, depressive states, and cognitive underperformance. Beta waves are further categorized into three subtypes. Low Beta (12–15 Hz), also termed beta-1, is linked to calm, focused, and introspective attention. Mid-range Beta (15–20 Hz), known as beta-2, is associated with increased cognitive effort, arousal, and performance-related anxiety. High Beta (18–40 Hz), or beta-3, is correlated with excessive mental stress, hypervigilance, paranoia, and sympathetic overactivation.

---

[3] ADHD, which stands for Attention-Deficit/Hyperactivity Disorder, is a neurodevelopmental disorder that typically emerges during childhood but can also persist into adulthood. It is characterized by difficulties in maintaining attention, impulsivity, and higher-than-normal levels of physical activity that are not appropriate for the context. The symptoms vary from person to person. Some individuals struggle to concentrate, are easily distracted, forget appointments, or lose things. Others exhibit impulsive behaviors, such as speaking without thinking, frequently interrupting others, or having trouble following rules. In many cases, there is also excessive restlessness, which may appear as a constant need to move, difficulty staying seated for long periods, or a persistent internal sense of agitation.

[4] Hyperarousal is a term used primarily in psychology and psychiatry to describe a state of heightened physiological and emotional alertness. It often occurs in response to stress or trauma and is commonly associated with conditions like PTSD (Post-Traumatic Stress Disorder), anxiety disorders, or ADHD. When someone experiences hyperarousal, their nervous system stays on high alert, even in the absence of danger. This can lead to symptoms

- Alpha Waves (8–12 Hz): Alpha rhythms occupy the frequency range between theta and beta waves and are most prominent during states of relaxed wakefulness and reduced sensory input. They play a pivotal role in promoting calmness and internal awareness. Alpha activity is typically elevated during states of mental idleness, such as daydreaming and relaxed alertness. Suppression of alpha waves may be associated with anxiety disorders, chronic stress, and insomnia. Optimal alpha activity contributes to a tranquil, yet wakeful, mental state conducive to restorative relaxation.

- Theta Waves (4–8 Hz): Theta oscillations are prevalent during light sleep, meditative states, and drowsiness, and are involved in introspective thought, creativity, and emotional regulation. Elevated theta activity is often observed in individuals with ADHD, depressive symptoms, impulsivity, and attentional deficits, while hypoactivity in this band can result in heightened anxiety, emotional dysregulation, and stress. In a balanced state, theta waves support intuitive cognition, emotional depth, creative ideation, and physiological restoration during sleep cycles.

By analyzing changes in the frequency and amplitude of these brainwaves in the PFC, it is possible to infer an individual's emotional and cognitive state [Nitsche et al., 2012; Al-Shargie et al., 2017].

3. Stress and Anxiety

Stress and anxiety are characterized by heightened activation of the amygdala and reduced regulation by the PFC. This imbalance results in specific patterns of electrical activity (e.g., increased beta waves). Moreover, the PFC mediates the subjective perception of stress, modulating the intensity of the stress response.

4. Cognitive Phases and Focus

During focus, the PFC exhibits greater coherence of gamma and beta waves, indicating high cognitive activity and inhibition of distractions. Specific activation patterns related to cognitive load are observed in complex cognitive tasks.

5. Techniques and Technologies for Analysis

EEG along with other techniques, such as functional near-infrared spectroscopy (fNIRS, which measures cerebral oxygenation), enables real-time recording and analysis of PFC activity. These non-invasive methods offer direct insights into cognitive and emotional processes.

The prefrontal cortex acts as a "hub" for emotional and cognitive regulation. The electrical signals from this region directly reflect underlying processes, making them sufficient to identify states such as stress, anxiety, attention, and cognitive activity. This is particularly useful in fields such as applied neuroscience, psychology, and brain-computer interface technology.

---

such as irritability, difficulty sleeping, exaggerated startle responses, racing thoughts, or a general feeling of being "on edge" all the time. It's as if the body and brain are stuck in a constant "fight or flight" mode. In short, hyperarousal means the body is overreacting to stimuli that wouldn't normally be considered threatening, making it hard to relax, concentrate, or feel safe.

Interaxon states on its website that many leading institutions use Muse's compact EEG system to access brainwave data in laboratories and real-world settings, including IBM, MIT, Yale, University of Toronto, Harvard, and NASA.

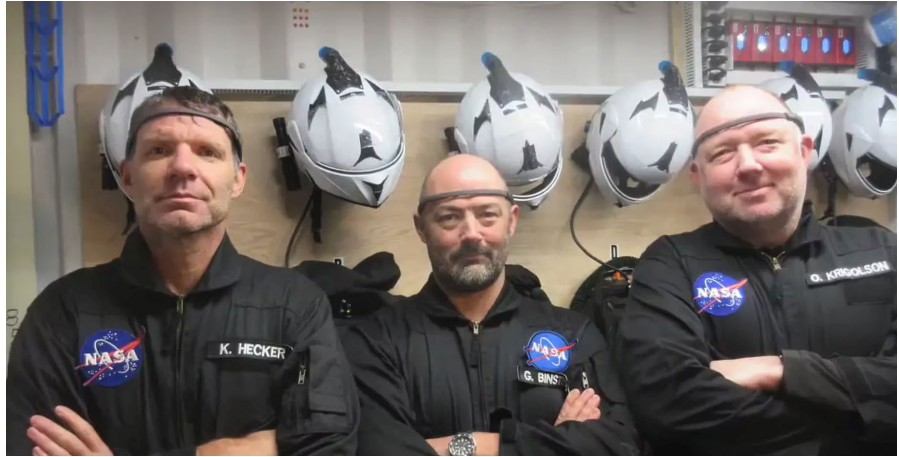

*Fig. 6 - Picture from the Interaxon web site.*

Muse has been pivotal in over 200 peer-reviewed studies focusing on cognitive performance, Attention Deficit Hyperactivity Disorder (ADHD), anxiety, Post-Traumatic Stress Disorder (PTSD), and more.

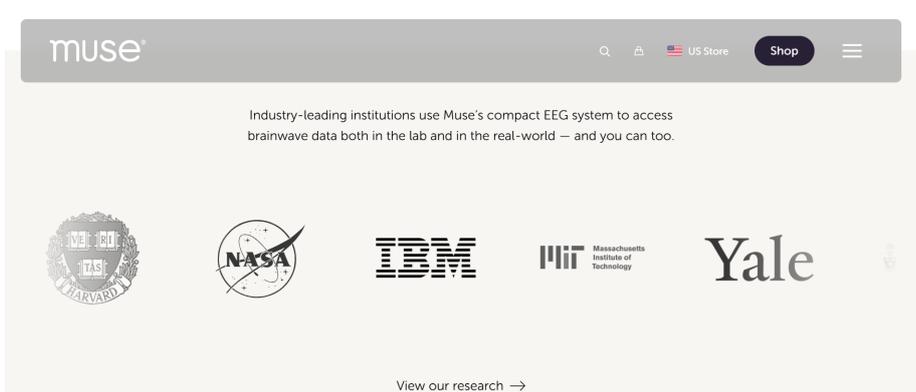

*Fig. 7 - Picture from the Interaxon web site.*

### 5. The data detection system and the analysis model.

This section details the operational methodology of the models utilized in Holytics by Sporthype. The data received from Muse is saved in files. The Muse device captures raw EEG (electroencephalography) signals, which reflect the brain's electrical activity and are measured in microvolts. Depending on the model, these signals are sampled at a high frequency, typically 256 Hz or 500 Hz. The device uses multiple electrodes positioned at specific locations on the scalp, such as TP9, AF7, AF8, and TP10, with a reference sensor placed at Fpz (see Fig. 4). From these raw EEG signals, it is possible to transform the time-domain signal into frequency-domain signal via spectral analysis. For example, delta waves (0.5–4 Hz) are linked to deep sleep, theta waves (4–8 Hz) to light sleep or meditative states, alpha waves (8–12 Hz) to calmness and relaxation, beta waves (12–30 Hz) to alertness and focused thinking, and gamma waves (30–100 Hz) to higher-level cognitive processing [Buzsáki, G., & Draguhn, A., 2004; Herrmann et al., 2004; Michal T Kucewicz et al., 2017]. This EEG data forms the core of Muse's neurofeedback capabilities.

The first operation that is performed is the control of the presence and consequent removal of any anomalous consecutive repetitions. Missing data (missing values) are also found and eliminated.

A low-pass filter is then applied to enhance signal analysis by reducing high-frequency noise—such as EMG activity, environmental interference, and electronic artifacts—thereby isolating low-frequency components and clarifying brain activity interpretation.

At this point, the Shannon Entropy (*entropy*)[5] is determined for each sensor of interest. Only the two frontal/anterior sensors are considered indispensable (left AF7, right AF8). Krigolson et al. (2017) in his study validated the use of the Muse EEG system for investigating event-related potentials (ERPs). The authors employed the frontal electrodes AF7 and AF8 to examine ERP components linked to cognitive processing, demonstrating their effectiveness in capturing neural signals associated with executive functions. The lateral electrodes TP9 and TP10 served as reference channels, thereby underscoring the central role of frontal sensors in the analysis of cognitively relevant brain activity. Hamedi et al. (2023) demonstrated that theta activity in the anterior prefrontal cortex, as recorded through frontal electrodes, is associated with memory and executive functions in patients with epilepsy. These findings suggest that frontal EEG sensors play a critical role in the monitoring and assessment of higher-order cognitive processes. Cannard et al. (2024): In this study, the authors compared the spectral characteristics of EEG signals obtained using both wearable and medical-grade devices. They observed that, with an appropriate reference montage, the frontal electrodes of the Muse headset (AF7 and AF8) can provide reliable measurements of frequency bands associated with well-being and meditation, underscoring their importance compared to lateral sensors. The entire column of the file relating to the sensor is extracted, the frequency of each value is counted, and it is normalized, obtaining a probability distribution. The entropy is calculated with the formula:

$$(1) \quad H(X) = -\sum p_i \, log_2(p_i)$$

where the $p_i$ represents the normalized probability associated with each value in the "AF" vector. If all entries in the "AF" are identical, the entropy equals zero, indicating an absence of uncertainty. Conversely, if the values are uniformly distributed, the entropy reaches its maximum. Higher entropy values correspond to greater uncertainty within the distribution.

The Welch method is applied to this [Welch P., 1967]. The Welch method applied to a millivolt EEG (electroencephalogram) dataset allows for the estimation of the power spectral density (PSD) of the signal in the frequency domain. From this estimate, the brain frequency bands are extracted. EEG is a non-stationary and noisy signal, so calculating the Fourier transform directly can lead to unstable results Welch's method produces a statistically consistent estimate of the power spectral density by averaging periodograms derived from partially overlapping segments of the time series, thereby reducing variance while preserving spectral resolution.

Splitting the EEG signal into overlapping time windows.

1. Applying a window function (e.g., Hann window) to reduce the leakage effect.
2. Calculating the Fourier transform for each window.

---

[5] Shannon entropy is a measure of uncertainty or information content introduced by Claude Shannon in 1948 as part of information theory. Shannon entropy quantifies the average amount of information produced by a random source of data.

3. Averaging the transformations to obtain the PSD, which indicates energy distribution in the frequency domain.

Once the PSD is obtained, the powers in the classic EEG bands can be extracted:

| Band | Frequency (Hz) |
|---|---|
| Delta | 0.5 - 4 Hz |
| Theta | 4 - 8 Hz |
| Alpha | 8 - 13 Hz |
| Beta low | 13 - 22 Hz |
| Beta high | 22 - 32 Hz |
| Gamma | 32 - 40 Hz |

*Tab. 1*

The operation is performed for each single row of the dataset, maintaining the original frequency generated by the EEG device.

The device sensors allow for determining the power of the two cerebral hemispheres (*AF7sx, AF8dx*), as the average of all the detected frequencies (average of alpha, low beta, high beta, gamma, theta) with the exclusion of delta, because they are typical of deep sleep.

The percentages of the individual frequencies are determined from the total power (all frequencies excluding delta). Theta and beta are taken into particular consideration (*theta%, beta%*).

The ratio between the high beta and total beta percentage values provides stress.

(2) $$stress = \frac{high\ beta\ \%}{beta\ \%}$$

With the frequency data obtained up to this point, it is possible to determine a series of KPIs. In particular, the following can be determined: *fog index, sharpness index, stress recovery index, cognitive load index, mental fatigue index, energy efficiency index, and energy consumption index*.

Isolating the elements *AF7sx, AF8dx, alpha%, theta%, beta%, stress, and entropy*, the flow index is determined.

6. **Case Study.**

Numerous professionals in the field report that one of the most frequently asked questions concerns the meaning and improvement of a golfer's mental state. Indeed, anyone who has played the sport has likely experienced a critical moment just before executing a shot: it is not the backswing nor the impact, but rather the five or six seconds that precede the swing. This brief time window can be the difference between a successful shot and a potentially compromising error. It is within this window that the so-called mental game comes into play. High-level athletes have developed the ability to manage their inner dialogue which, if left unchecked, can undermine performance—or, conversely, facilitate access to the state known as flow. In this context, the mental game in golf can be defined as "the ability to unlock unconscious potential through a conscious and strategic management of the rational component" (Giorgio De Pieri). While this concept may initially seem complex to beginners, numerous neuroscientific studies indicate that the prefrontal cortex—the seat of rationality and voluntary cognitive control—tends to exhibit excessive activity during moments of pressure or uncertainty. To access peak motor efficiency, such as the flow state, it is therefore essential to reduce interference from this brain region, allowing for a temporary disengagement of conscious control in favor of automatism and unconscious resources.

EEG recordings were conducted using the Muse 2 EEG device on two professional male golf coaches in their forties. A total of 84 recordings were collected.

A detection protocol was employed that included both visualization exercises and focused attention meditation, as well as athletic practices involving actual shot execution. It is important to note that, even prior to shot execution, golfers routinely perform visualization as part of their preparatory routine. Following each technical execution, participants were asked to complete a questionnaire comprising ten items concerning their subjective experience. The items were rated on a 5-point Likert scale, where 1 corresponded to "Strongly disagree" and 5 to "Strongly agree."

- Q1: During the activity, I experienced a sense of calm and inner stability.
- Q2: I had a clear perception of my goals without feeling overwhelmed by pressure.
- Q3: I felt in perfect harmony with my body and movements.
- Q4: I perceived a balance between concentration and relaxation.
- Q5: I was in control of my mind, free from distracting or negative thoughts.
- Q6: Time seemed to pass either faster or slower.
- Q7: I maintained steady focus effortlessly, as if everything was flowing naturally.
- Q8: I experienced a sense of well-being and serenity during the activity, regardless of performance.
- Q9: I did not feel anxiety or stress, but rather a sense of harmony and connection with the activity.
- Q10: The activity left me with a deep sense of satisfaction and mental clarity.

Some recordings were taken during the execution of an entire golf hole, while others focused on individual shots.

The test sessions were conducted during regular training activities, both in indoor and outdoor training centers, in environments not shielded from potential external disturbances. Some EEG recordings were collected using a professional green simulator, while others took place on traditional practice greens with real putting surfaces. In the latter case, the putts were performed on an actual green.

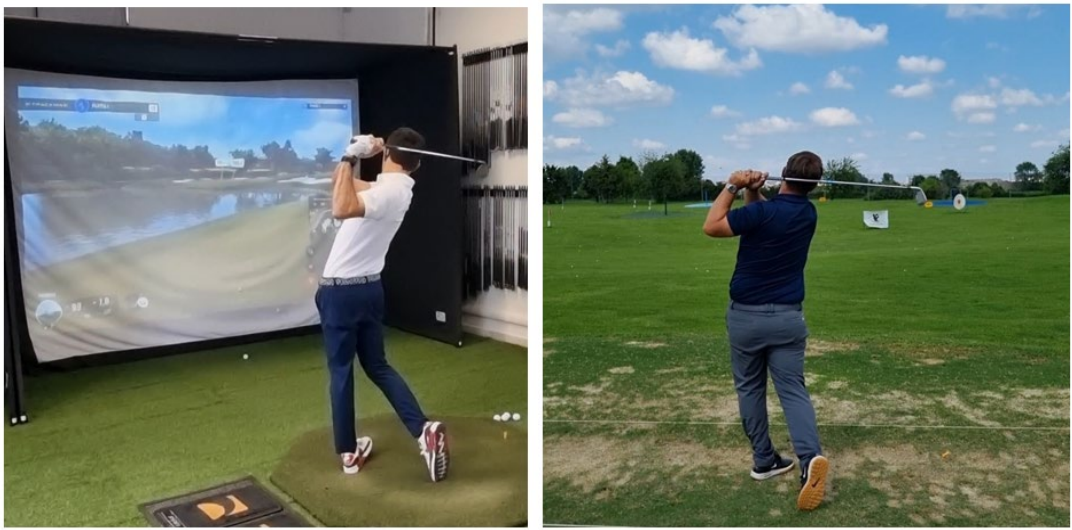

*Fig. 8 - Tests.*

Focused Attention Meditation (FAM) is a meditative practice in which the participant intentionally directs attention to a specific object. In this study, participants engaged in focused meditation by concentrating on a self-selected anchor point. This technique is intended to train the mind to identify and reduce distractions, thereby enhancing concentration and attentional control.

In a typical golf hole, play begins with a drive from the tee, where the player attempts to maximize distance toward the target, ideally landing the ball on the fairway. This is followed by one or more approach shots aimed at positioning the ball on the green. Once on the green, the final stage involves putting, with the goal of accurately rolling the ball into the hole.

In our experiment, we adopted this natural sequence of play to examine neural activity across the different phases, comparing the cognitive demands and EEG patterns during the driving phase versus the putting phase. Some measurements were taken over the course of an entire hole, while others focused on single actions.

EEG signals were pre-processed to remove artifacts and noise using filtering techniques and single-subject Independent Component Analysis (ICA). Machine learning algorithms, including k-Nearest Neighbors (k-NN), Principal Component Analysis (PCA), and K-means clustering, were then applied to identify and classify the primary patterns of brain activity. The "flow" pattern was validated through comparison with data from the professional coach, assessing the reproducibility and stability of the signal across sessions.

Throughout the different sessions, the coach performed the same familiar exercises and movements, ensuring accuracy and consistency in athletic execution, with no uncertainty regarding technique.

Visualization exercises yielded a slightly lower Flow State Index (FSI) compared to actual physical performance. The athlete's self-reported experience, as reflected in the questionnaire responses, closely matched the FSI calculated by the Holytics algorithm, with a discrepancy of approximately 9%, which we consider to fall within a physiological range.

|  | Flow average | Survey average | Ratio average |
|---|---|---|---|
| execution | 0,72 | 0,75 | 9% |
| visualization | 0,66 |  |  |
|  | 0,70 | 0,75 | 9% |

Tab. 2

We also consider it more common for individuals to overestimate their mental performance subjectively, rather than to underestimate it relative to objectively measured data. Indeed, such underestimations accounted for only 1 out of every 5 cases.

Regarding location, results were mixed: motor execution was found to be more effective in indoor settings, whereas visualization appeared more effective outdoors.

|  | Location | Flow average |
|---|---|---|
| execution | indoor | 0,77 |
|  | outdoor | 0,66 |
| visualization | indoor | 0,65 |
|  | outdoor | 0,83 |

Tab. 3

Among all athletic gestures, the drive was associated with the highest FSI, while approach shots and putts (i.e., the final stages of the hole) produced the lowest values.

|                    | Flow average |
|--------------------|--------------|
| Drive              | 0,77         |
| Fairway Shot       | 0,70         |
| Approach Shot      | 0,65         |
| Chip Shot or Putt  | 0,65         |

*Tab. 4*

It is worth noting that some very high flow values (greater than 0.9) were recorded. In these cases, when perception monitoring is present, a convergence is observed that indicates the participants' awareness of having performed the putt flawlessly, where "flawlessly" refers to a maximal correspondence between expectations and execution.

### 7. Activities and exercises for Flow State Training (FST).

There are several activities that can be performed to train the mind to reach and maintain an optimal level of flow. These exercises are recognized as effective for this specific purpose. Below are some of the most used:

1. Diaphragmatic Breathing: also referred to as abdominal or belly breathing, this method engages the diaphragm to promote parasympathetic nervous system activation. It has been shown to reduce physiological stress markers and enhance cognitive clarity, supporting the neurophysiological conditions conducive to flow.
2. Box Breathing (4-4-4-4): a structured breathing technique involving equal intervals of inhalation, breath-holding, exhalation, and another breath-holding phase (each lasting four seconds). This approach has been demonstrated to stabilize autonomic function and enhance focus, particularly under conditions of pressure.
3. Mindfulness practice: it involves maintaining a non-judgmental, moment-to-moment awareness of present experience. Empirical studies have linked regular mindfulness practice with improved attentional control, emotional regulation, and reduced mind-wandering—all of which facilitate the onset of flow.
4. Focused Attention Meditation: this practice requires sustained attention on a chosen object (e.g., the breath or a mantra). Repeated engagement in focused attention meditation enhances executive control mechanisms, thereby improving the ability to maintain concentration over extended periods.
5. Guided Visualization: a cognitive simulation technique in which individuals are led through vivid mental imagery of desired outcomes or performance scenarios. Guided visualization has been shown to prime motor and cognitive circuits, reduce anticipatory anxiety, and increase task-specific self-efficacy.
6. Mental Rehearsal: closely related to guided visualization, mental rehearsal involves the cognitive representation and internal repetition of task execution. This process supports procedural learning and mental preparedness, providing seamless real-time performance.
7. Instrumental and binaural music: listening to instrumental or binaural beat-based auditory stimuli can influence cortical activity and support flow induction. Binaural beats, in particular, have been linked to brainwave training, which may facilitate states of deep concentration and relaxation.

8. Brainwave frequency stimulation (40, 10, 7 Hz): specific auditory or neurofeedback-based interventions targeting gamma (40 Hz), alpha (10 Hz), and theta (7 Hz) frequencies are associated with enhanced cognitive processing, alert relaxation, and creativity, respectively. Modulation of these frequencies may support transitions into and maintenance of the flow state.
9. Light Stretching and Yoga-based movements: engaging in low-intensity physical activity such as stretching or yoga has been correlated with reductions in somatic tension and improvements in interoceptive awareness. These physiological shifts contribute to a more integrated mind-body state, thereby facilitating the emergence of flow.
10. Neuro-Linguistic Programming (NLP): mental anchoring involves associating a specific physical or cognitive cue with a desired psychological state. Over time, this associative learning process can enable rapid induction of flow-related mental states through intentional cue activation.
11. Pomodoro Technique (25-5): this time management strategy alternates 25-minute periods of focused work with 5-minute rest intervals. The Pomodoro Technique has been associated with enhanced sustained attention, reduced cognitive fatigue, and improved task engagement, all of which are compatible with the cognitive conditions of flow.

Standard usage protocols are available for each of these exercises, and they are easy to follow. In any case, the support of a psychologist or mental coach can certainly be beneficial. Performing exercises of this nature is an essential complement to the measurement of flow state as described in the previous section. Ongoing evaluation of the flow state over time may include moments when these exercises become helpful or even indispensable. Mental training is just as important as physical training, and its measurement represents an element of objectivity required in professional practice.

Some of these exercises are preparatory for self-hypnosis activities. Flow is a state that occurs spontaneously, but it is not always easy to get into it: one needs mental clarity, absence of distractions, and a certain optimal psychophysical state. Self-hypnosis is an effective tool to create just that fertile ground. It increases selective focus (during self-hypnosis you learn to isolate your attention to a single thought or goal, ignore irrelevant stimuli), shuts down the inner critic (self-hypnosis reduces the activity of the dorsolateral prefrontal cortex, which is linked to self-criticism and judgement), regulates the emotional and physiological state, and sets a powerful mental intention. During self-hypnosis, individuals can visualize themselves going into flow.

## 8. Leveraging Artificial Intelligence to Analyze Outputs and Deliver Actionable Recommendations.

Traditional neurophysiological signal analysis has long centered on identifying strong signals – clear, high-amplitude brainwave events that stand out against the noisy background of EEG data. Classic brain-computer interface (BCI) research exemplified this focus: for instance, the P300 event-related potential and steady-state visual evoked potentials (SSVEPs) are robust signals that can be reliably induced and detected with accuracies often above 90% [Farwell and Donchin, 1988]. Early breakthroughs leveraged such strong signals (e.g., Farwell and Donchin's 1988 communication system using the P300) because their large amplitude and distinct patterns could be extracted by simple averaging or threshold-based methods. By contrast, "weak signals" – weak patterns associated with endogenous cognitive states – have been historically elusive. EEG recordings suffer extremely low signal-to-noise ratios; a cortical signal on the order of microvolts is heavily attenuated by the skull and mingled with other brain activity. As a result, small shifts in oscillatory power or fleeting micro-patterns of brain activity are at present technology lost in the noise [Dong et al., 2021]. These limitations of classical EEG analysis, which relies on averaging out noise and focusing on time-locked high-amplitude features, highlight a critical gap.

AI and advanced statistical methods have rapidly expanded the toolkit for decoding cognitive states from brainwaves, bringing new capabilities to both neuroscience research and sports performance monitoring. Improvements in mobile EEG technology and machine learning algorithms, it is now feasible to monitor an athlete's brain activity in real-world training environments (e.g. during tennis or golf practice) without impeding movement [Gupta et al., 2024].

AI-powered EEG analysis is revolutionizing attention monitoring and neurofeedback, enabling real-time detection of cognitive states, such as focus and distraction, in athletes. Through machine learning, EEG data is analyzed to identify when an individual's concentration is slipping, allowing for corrective actions like neurofeedback training [Perrey, 2022]. Wearable devices are the physical instrumentation that might help in reaching such a results.

AI in brainwave monitoring is being used to predict performance success or failure by analyzing pre-performance neural patterns. Machine learning models can identify subtle EEG signatures that correlate with an athlete's mental readiness and likelihood of optimal performance. For example, in a 2024 e-sports study, researchers found that elevated beta-band activity in the parietal cortex predicted success, while higher frontal alpha and gamma power indicated likely losses, achieving an 80% accuracy in predicting outcomes [Minami et al., 2024]. This approach highlights the potential for subtle cognitive-state markers to act as early indicators of performance. In addition, AI in EEG analysis has advanced beyond basic classifiers to incorporate deep learning architectures, such as deep neural networks (DNNs) and transformers, which analyze complex spatiotemporal patterns in raw EEG data. These models excel in detecting subtle neural signatures and have been particularly successful in fields like mental health diagnostics [Liu and Zhao, 2025]. In sports, AI systems now combine EEG with other data streams, such as video footage, to assess both neural and physical performance simultaneously. For example, a 2024 study used a transformer-based model to analyze EEG and video data, significantly improving accuracy in recognizing movement patterns and performance states [Sun, 2024].

The aforementioned results are suggesting that AI-driven brainwave monitoring is set to revolutionize sports training by enhancing real-time cognitive state assessments and optimizing performance. Future applications will go beyond detection, offering AI-guided EEG feedback to tailor training routines based on an athlete's brain signals. Wearable EEG devices, like headbands or sports caps, will allow constant monitoring, enabling AI to adjust exercises according to the athlete's mental engagement. If focus drops, AI might suggest a new activity or mindfulness break, while high concentration could lead to increased difficulty to maximize training. This integration of cognitive metrics with physical measures will ensure athletes train at the optimal mental load. AI will also enhance learning and skill acquisition by identifying peak receptivity moments, when athletes are most ready to learn new skills. AI can personalize training methods based on an athlete's cognitive profile, such as their preference for visual or verbal learning strategies, and detect early signs of mental fatigue to adjust training accordingly. The fusion of EEG with other modalities like eye-tracking and motion sensors will further enrich performance assessments, providing a comprehensive view of mental and physical states. In the future, AI will also aid in mental wellness, guiding athletes through personalized neurofeedback sessions to manage stress, improve focus, and optimize recovery. Overall, AI-powered EEG systems will drive individualized, data-driven performance optimization, ensuring athletes reach their full cognitive and physical potential.

Future applications will go beyond detection, offering AI-guided EEG feedback to tailor training routines based on an athlete's brain signals. Wearable EEG devices, like headbands or sports caps, will allow constant monitoring, enabling AI to adjust exercises according to the athlete's mental engagement. If focus drops, AI might suggest a new activity or mindfulness break, while high concentration could lead to increased difficulty to maximize training [Pei et al., 2022]. This integration of cognitive metrics with physical measures will ensure athletes train at the optimal mental load. AI will also enhance learning and skill acquisition by identifying peak receptivity moments, when athletes are most ready to learn new skills.

## 9. Conclusions.

The primary field targeted by this study is sports. However, not all tests conducted for empirical evaluation were performed for sports-related purposes, considering that the potential applications are so numerous that listing them all would be impractical. As an example, any activity requiring a flow state, as described in this study, can be found in a wide range of everyday activities. Furthermore, it should be noted that the flow state is closely linked to safety criteria in task execution and, in some ways, serves as a counterbalance to anxiety and, even more so, to fear.

The article highlighted how professional golfers frequently experience the flow state during the execution of their shots. This mental state, characterized by heightened concentration and a relaxed focus, proves to be crucial for optimal stress management and a clear, goal-oriented vision.

From a neurophysiological perspective, flow is associated with a temporary deactivation or disinhibition of the dorsolateral prefrontal cortex (a phenomenon known as transient hypofrontality), which is responsible for critical thinking, self-awareness, and judgment. This condition allows for a reduction in internal dialogue and distracting thoughts, thereby promoting greater cognitive and motor efficiency. At the same time, low-frequency brain waves, particularly alpha and theta waves, are commonly associated with relaxed alertness and access to deep cognitive resources.

In summary, the flow state represents an optimal condition for athletic performance, in which body and mind operate in perfect synchrony, minimizing cognitive interference and maximizing the effectiveness of action.

The results of the study revealed a significant alignment between EEG data and the subjective experiences reported in the questionnaires, confirming the feasibility of detecting the flow state through prefrontal cortex activity. Furthermore, the psychological exercises included in the protocol demonstrated a tangible positive effect in enhancing flow during athletic performance.

The integration of EEG measurements and questionnaires served as a robust control test for validating the flow estimation algorithm. This approach allows for the objectification of the individual's subjective experience, overcoming limitations due to a lack of self-awareness or self-knowledge, which can compromise the reliability of self-assessments. In this way, athletes can rely on an immediate and accurate estimation of their flow state, enabling real-time monitoring of focus and calm, and allowing timely intervention through mental exercises when the flow state is suboptimal.

## 10. Acknowledgements.

The authors would like to thank *Emanuele Bianco* (Coach of Edoardo Molinari Development Center) and *Alberto Campanile* (Coach of Golfus Performance Center) for the helpful discussions and valuable insights that contributed to the development of this work, especially during their training sessions for data collection. We thank *SPORTHYPE* for its financial and technical support for the fieldwork. This work was supported by PRIN PNNR grant (PIAL_PNRR_PRIN22_23_01) to Pia L.

___

## 11. References.

Antonini Philippe, R., et al. (2022). "Achieving Flow: An Exploratory Investigation of Elite College Athletes and Musicians." Psychology of Sport and Exercise.

Avvenuti, G., S. Varanese, e A. D'Errico. 2022. "A Retrospective Analysis of Three Focused Attention Meditation Techniques: Mantra, Breath, and External-Point Meditation." Frontiers in Psychology. Accessed March 20, 2025. https://pubmed.ncbi.nlm.nih.gov/35386478/.

Balconi M. et al., Be creative to innovate! EEG correlates of group decision-making in managers, MDPI Sustainability 2024, 16, 2175, (2024).

Bandler, R., Grinder, J. (1979), "Frogs into Princes: Neuro Linguistic Programming", Real People Press.

Bauer CCC, Rozenkrantz L, Caballero C, Nieto-Castanon A, Scherer E, West MR, Mrazek M, Phillips DT, Gabrieli JDE, Whitfield-Gabrieli S. Mindfulness training preserves sustained attention and resting state anticorrelation between default-mode network and dorsolateral prefrontal cortex: A randomized controlled trial. Hum Brain Mapp. 2020 Dec 15;41(18):5356-5369. doi: 10.1002/hbm.25197. Epub 2020 Sep 24. PMID: 32969562; PMCID: PMC7670646.

Bird J.J. et al., A Study on Mental State Classification usign EEG-based Brain-Machine Interface, 2018 International Conference on Intelligent Systems, (2018).

Braboszcz, C., et al. (2024). "Report from a Tibetan Monastery: EEG neural correlates of concentrative and analytical meditation." PMC.

Brown, R. P., & Gerbarg, P. L. (2005). Sudarshan Kriya Yogic breathing in the treatment of stress, anxiety, and depression: Part II—clinical applications and guidelines. Journal of Alternative and Complementary Medicine.

Cannard C. et al., Electroencepohalography correlates of well-being using a low-cost wearble system, Frontiers in Human Sciences, v 15, art 745135, (2021).

Chaudhary M. et al., Understanding Brain Dynamics for Color Perception using Wearble EEG headband, arXiv 2008.07092v1, (2020).

Chen, H., Liu, C., Zhou, F., Cao, X. Y., Wu, K., Chen, Y. L., ... & Chiou, W. K. (2022). Focused-Attention Meditation Improves Flow, Communication Skills, and Safety Attitudes of Surgeons. International Journal of Environmental Research and Public Health, 19(9), 5292. https://doi.org/10.3390/ijerph19095292::contentReference[oaicite:7]{index=7}.

Cannard, J., et al. (2024). EEG correlates of well-being and wearable EEG. Academia.edu.

Cheron, G., et al. (2016). Brain oscillations in sport: Toward EEG biomarkers of performance. Frontiers in Psychology, 7, 246.

Cheruvu, R. (2018). "The Neuroscience of Flow." Journal of Cognitive Enhancement.

Cirillo, F. (2006). The Pomodoro Technique. Pomodoro Technique Official Website.

Corrado, S. (2024). Improving Mental Skills in Precision Sports by Using Neurofeedback Training: A Narrative Review. Sports, 12(3), 70.

Csíkszentmihályi, M. (1975). Beyond Boredom and Anxiety: Experiencing Flow in Work and Play. San Francisco, CA: Jossey-Bass. ISBN 978-0875892610.

Csíkszentmihályi, M. (1990). Flow: The Psychology of Optimal Experience. New York: Harper & Row. ISBN 978-0060920432.

Csíkszentmihályi, M. (1996). Creativity: Flow and the Psychology of Discovery and Invention. New York: Harper Perennial. ISBN 978-0060928209.

Csíkszentmihályi, M. (1998). Finding Flow: The Psychology of Engagement with Everyday Life. New York: Basic Books. ISBN 978-0465024117.

Csíkszentmihályi, M. (2003). Good Business: Leadership, Flow, and the Making of Meaning. New York: Penguin Books. ISBN 978-0142004098.

Csíkszentmihályi, M. (2014). Flow and the Foundations of Positive Psychology. Springer Dordrecht. ISBN 978-94-017-9087-1.

Dietrich, A. (2003). Functional Neuroanatomy of Altered States of Consciousness: The Transient Hypofrontality Hypothesis. Consciousness and Cognition, 12(2), 231-256.

Dietrich, A. (2004). "Neurocognitive mechanisms underlying the experience of flow." Consciousness and Cognition.

Dilts, R. (1990). Changing Belief Systems with NLP. [Meta Publications].

Dilts, R., Hallbom, T., & Smith, S. (1990). Beliefs: Pathways to health and well-being. Metamorphous Press..

Dong, H. W., et al. (2021). Detection of mind wandering using EEG: Within and across individuals. PLOS ONE, 16(5), e0251490.

Dove, H. W. (1839), scoperta dei toni binaurali.

Driskell, J. E., Copper, C., & Moran, A. (1994). Does mental practice enhance performance?. Journal of Applied Psychology.

Field, T. (2011). Yoga clinical research review. Complementary Therapies in Clinical Practice.

Getting into a 'Flow' State: A Systematic Review of Flow Experience in Neurological Diseases (2021). Journal of NeuroEngineering and Rehabilitation.

Ghosh K. et al., Mindfulness Using a Wearble brain Sensing Device fir Healt Care Professional During a Pandemic: a pilot program, J of primary care & community healthj v 14: 1-10, (2023).

Girivirya S., Analysis of Mindfulness Practices Using Elctroencephalogram (EEG) Interaxon Muse Heandband Against the Concept of Self-Acceptance of Poststroke Clients, Influence, vol 5, np 2, 2023, (2023).

Glavas C. et al., Evaluation of the User Adaptaion in BCI Game Environment, Appl Sci. 2022, 12, 12722, (2022).

Gold, J., & Ciorciari, J. (2020). "A Review on the Role of the Neuroscience of Flow States in the Modern World." PMC.

Gruzelier, J. H. (2009). A theory of alpha/theta neurofeedback, creative performance enhancement, long distance functional connectivity and psychological integration. Cognitive Processing, 10(1), 101-109. https://doi.org/10.1007/s10339-008-0248-5:contentReference[oaicite:15]{index=15}.

Gu, J., et al. (2015). How do mindfulness-based cognitive therapy and mindfulness-based stress reduction improve mental health and wellbeing? A systematic review and meta-analysis of mediation studies. Clinical Psychology Review.

Guo, J.-H., et al. (2025). Enhancing shooting performance and cognitive engagement in virtual reality environments through brief meditation training. Scientific Reports, 15, 16289.

Gupta, E., Chen, C.-Y., & Sivakumar, R. (2024). Do Weak Brain Signals Get Amplified When Strong Brain Signals are Evoked? In Proceedings of the IEEE PerCom Workshops (pp. 1–7).

Hamedi, N., García-Salinas, J. S., Berry, B. M., Worrell, G. A., & Kucewicz, M. T. (2025). Anterior prefrontal EEG theta activities indicate memory and executive functions in patients with epilepsy.

Harris, D. J., Vine, S. J., & Wilson, M. R. (2017). Is Flow Really Effortless? The Role of Effort and Monitoring in Flow States. Human Movement Science, 57, 153-162.

Herman K. et al., Emotional Well-Being in Urban Wilderness: Assessing states of calmness and alertness in informal green spaxces (IGSs) with Muse-Portable EEG band, Sustainability 2021, 13, 2212, (2021).

Herrmann, C. S. (2001). Human EEG responses to 1–100 Hz flicker: resonance phenomena in visual cortex and their potential correlation to cognitive phenomena. Experimental Brain Research.

Hiltner et al., An overview of EEG artifacts, perceptio and task-related brain activity, Applied Neuroscience WS2020/2021, (2021).

Isham, A., & Jackson, T. (2022). "Finding Flow: Exploring the Potential for Sustainable Fulfilment." Sustainability Science.

Jackson, S. A., & Csíkszentmihályi, M. (1999). Flow in Sports: The Keys to Optimal Experiences and Performances. Champaign, IL: Human Kinetics Publishers. ISBN 978-0880118767.

Jaison F. et al., EEG-based brain-machine interface for categorizing cognitive sentimental emotions, Multidisciplinary Science Journal aug 10, 2023, (2023).

Java Sutawan IBK et al., calm-mind percentage measured by a wearble electroencephalogram (EEG) eandband correlates with stress level in anesthesia residents: a prospective, observational study, Bali J Anesthesiol 2024;8:69-73, (2024).

Karmanews (2024), "La musica binaurale", intervista alla dott.ssa Mary Hoffer, Spa Music Institute.

Karmanews (2024), "La musica binaurale" (spiegazione degli effetti delle frequenze cerebrali).

Katahira K, Yamazaki Y, Yamaoka C, Ozaki H, Nakagawa S and Nagata N (2018) EEG Correlates of the Flow State: A Combination of Increased Frontal Theta and Moderate Frontocentral Alpha Rhythm in the Mental Arithmetic Task. Front. Psychol. 9:300. doi: 10.3389/fpsyg.2018.00300.

Kaufman, K. A., Glass, C. R., & Pineau, T. R. (2018). Mindful Sport Performance Enhancement (MSPE): Effects on flow state and mental health among athletes. Journal of Clinical Sport Psychology, 12(3), 234-250. https://doi.org/10.1123/jcsp.2017-0020:contentReference[oaicite:5]{index=5}.

Klimesch, W. (1999). "EEG alpha and theta oscillations reflect cognitive and memory performance: a review and analysis." Brain Research Reviews.

Krigolson and colleagues (2017), titled "Choosing MUSE: Validation of a Low-Cost, Portable EEG System for ERP Research"; this study, published in Frontiers in Neuroscience, evaluates the effectiveness of the Muse portable EEG system in ERP [event-related potential] research).

Krigolson O.E. et al., Choosing MUSE: validation of a Low-Cost, Portable EEG system for ERP Research, Front. Neurosci. 11:109. doi: 10.3389/fnins.2017.00109, (2017).

Krigolson O.E. et al., Using Muse: Rapid Mobile Assessment of Brain Performance, Front. Neurosci. 15:634147. doi: 10.3389/fiins.2021.634147, (2021).

Lane, J. D., Kasian, S. J., Owens, J. E., & Marsh, G. R. (1998). Binaural auditory beats affect vigilance performance and mood. Physiology & Behavior.

Lane, J. D., Kasian, S. J., Owens, J. E., & Marsh, G. R. (1998). Binaural auditory beats affect vigilance performance and mood. Physiology & Behavior, 63(2), 249-252. https://doi.org/10.1016/S0031-9384(97)00436-8:contentReference[oaicite:13]{index=13}.

Lang, P. J. (1979), "A bio-informational theory of emotional imagery." Psychophysiology, 16, 495-512.

Lardone, A., M. Liparoti, P. Sorrentino, R. Rucco, F. Jacini, C. Cavaliere, e R. Formisano. 2020. "Focused Attention Meditation Training Modifies Neural Activity and Attention: Longitudinal EEG Data in Non-Meditators." NeuroImage. Accessed March 20, 2025. https://pubmed.ncbi.nlm.nih.gov/32064537/.

Lazrou I. et al., Eliciting brain waves of people with cognitive impairment during meditation exercises using portable electroencephalography in a smart-home environment: a pilot study, Frontiers in Aging Neuroscience 30 may 2023, (2023).

Lee, J., et al. (2020). Effects of Diaphragmatic Breathing on Health: A Narrative Review. MDPI.

Li, Y., et al. (2021). "Longitudinal effects of meditation on brain resting-state functional connectivity." Scientific Reports.

Liu, Z., & Zhao, J. (2025). Leveraging deep learning for robust EEG analysis in mental health monitoring. Frontiers in Neuroinformatics, 18, Art. 1494970.

Lutz, A., Slagter, H. A., Dunne, J. D., & Davidson, R. J. (2008). "Attention regulation and monitoring in meditation." Trends in Cognitive Sciences.

Lutz, A., Slagter, H. A., Dunne, J. D., & Davidson, R. J. (2008). Attention regulation and monitoring in meditation. Trends in Cognitive Sciences.

Ma, X., Yue, Z. Q., Gong, Z. Q., Zhang, H., Duan, N. Y., Shi, Y. T., ... & Li, Y. F. (2017). The effect of diaphragmatic breathing on attention, negative affect and stress in healthy adults. Frontiers in Psychology, 8, 874.

Majid M. et al., PROPER: persoanlity recognition based on public speaking using electroencephalography recording, IEEE Access volume 4, 2016, (2016).

Malenka Robert C., Bear Mark F. (2004), LTP and LTD: An Embarrassment of Riches, doi: 10.1016/j.neuron.2004.09.012, Elsevier.

Mansi S.A. et al., EEG measurements-based study for evaluating ecoustic human perception: a pilot study, ACTA IMEKO june 2024, v 12, n. 2, 1-9, (2024).

Martin, K. A., Moritz, S. E., & Hall, C. (1999), "Imagery use in sport: A literature review and applied model." The Sport Psychologist, 13, 245-268.

Massimini, F., & Carli, M. (1988). "The Systematic Assessment of Flow in Daily Experience." In Optimal Experience.

Milena Girotti, Samantha M. Adler, Sarah E. Bulin, Elizabeth A. Fucich, Denisse Paredes, David A. Morilak (2018). Prefrontal cortex executive processes affected by stress in health and disease, Progress in Neuro-Psychopharmacology and Biological Psychiatry, Volume 85, Pages 161-179, ISSN 0278-5846,.

Minami, S., Watanabe, K., Saijo, N., & Kashino, M. (2024). Best brain conditions for winning an esports competition: EEG amplitude in the frontal and parietal cortices associated with competition results. In Proceedings of the 46th Annual Conference of the Cognitive Science Society (pp. 1024–1030).

Moran, A. (2012). Sport and exercise psychology: A critical introduction. Routledge..

Nakamura, J., & Csíkszentmihályi, M. (2002). The Concept of Flow. In C. R. Snyder & S. J. Lopez (Eds.), Handbook of Positive Psychology (pp. 89-105). Oxford: Oxford University Press. ISBN 978-0195135336.

Negrón-Oyarzo I, Aboitiz F, Fuentealba P. Impaired Functional Connectivity in the Prefrontal Cortex: A Mechanism for Chronic Stress-Induced Neuropsychiatric Disorders. Neural Plast. 2016;2016:7539065. doi: 10.1155/2016/7539065. Epub 2016 Jan 19. PMID: 26904302; PMCID: PMC4745936.

Paylo, M. J., & Zoldan, C. A. (2013). The effects of visualization and guided imagery in sports. Texas State University. Retrieved from https://digital.library.txst.edu/handle/10877/4863:contentReference[oaicite:9]{index=9}.

Pei, X., et al. (2022). A simultaneous electroencephalography and eye-tracking dataset in elite athletes during alertness and concentration tasks. Scientific Data, 9, 465.

Perrey, S. (2022). Training Monitoring in Sports: It Is Time to Embrace Cognitive Demand. Sports, 10(5), 70.

Priyanka A. Abhang, Bharti W. Gawali, Suresh C. Mehrotra (2016) "Introduction to EEG- and Speech-Based Emotion Recognition", chapter 3 Technical Aspects of Brain Rythms and Speech Parameters.

Prapas G. et al., Mind the Move: Developing a Brain-Computer Interface Game with Left-Right Motor Imagery, Information 2023, 14, 354, (2023).

Reid, M. R., J. Lutz, e K. A. Garrison. 2021. "Evaluating the Feasibility of a Consumer-Grade Wearable EEG Headband to Aid Assessment of State and Trait Mindfulness." Mindfulness Journal. Accessed March 20, 2025. https://pubmed.ncbi.nlm.nih.gov/34061998/.

Sharma K. et al., A retrospective analysis if three focused attention meditation techniques: mantra, breath, and external-point meditation, Cureus 14(3): e23589, (2022).

Sharma, M., et al. (2021). Effects of diaphragmatic deep breathing exercises on prehypertensive or hypertensive individuals. ScienceDirect.

Sheikh, A. A., Sheikh, K. S., Moleski, L. M. (1994), "Improving imaging abilities." In Sheikh and Korn (Ed.), Imagery in sports and physical performance. Baywood Publishing Company.

Shernoff, D. J., et al. (2003). "Student Engagement in High School Classrooms from the Perspective of Flow Theory." School Psychology Quarterly.

Shonin, E., Van Gordon, W., & Griffiths, M. D. (2014). Mindfulness-based interventions: Towards mindful clinical integration. Frontiers in Psychology.

Sidelinger L. et al., Day-to-day individual alpha frequency variability measured by a mibile EEG device related to anxiety, Eur J Neurosci. 2023;57:1815-1833, (2023).

Simon, M. 2023. "I Tried These Brain-Tracking Headphones That Claim to Improve Focus." WIRED Magazine. Accessed March 20, 2025. https://www.wired.com/story/this-brain-tracking-device-wants-to-help-you-work-smarter/.

Sun, Q. (2024). EEG-powered cerebral transformer for athletic performance. Frontiers in Neurorobotics, 18, Art. 1499734.

Sutanto E. et al., Implementation of Closing Eyes Detection with ear sensor of Muse EEG headband using support vector machine learning, Int J of Intelligent Engineering & Systems, (2022).

Sànchez-Cifo M.A. et al., MuseStudio: Brain Activity Sata management Library for Low-Cost EEG Device, Appl Sci. 2021, 11, 7644, (2021).

Tian K., Muse Headband: potential communication tool for locked-in people, Mechanical Engineering research vol 8 n.1 2018, (2018).

Tse, D. C. K., Nakamura, J., & Csikszentmihalyi, M. (2022). "Flow Experiences Across Adulthood: Preliminary Findings on the Continuity Hypothesis." Psychology and Aging.

Ulrich, B., Keller, J., & Grön, G. (2016). Neural Signatures of Experiential Involvement in Tetris. Neuropsychologia, 90, 165-173.

Verywell Mind. (n.d.). Box Breathing Techniques and Benefits. Retrieved from https://www.verywellmind.com/the-benefits-and-steps-of-box-breathing-4159900:contentReference[oaicite:3]{index=3}.

Wang, C. K. J., et al. (2016). Brain Mechanisms of the Flow State: A Review. Frontiers in Human Neuroscience, 10(43), 1-14.

Wang, Y.-K., Jung, T.-P., & Lin, C.-T. (2015). EEG-based attention tracking during distracted driving. IEEE Transactions on Neural Systems and Rehabilitation Engineering, 23(6), 1085–1094.

Welch P. (1967). The use of fast Fourier transform for the estimation of power spectra: a method based on time averaging over short, modified periodograms. IEEE Trans. Audio Electroacoust. 15, 70–3.

Wolfgang Klimesch, 1999, EEG alpha and theta oscillations reflect cognitive and memory performance: a review and analysis, Brain Research Reviews, Volume 29, Issues 2–3.

Woolf, M. 2023. "Can a £500 Gadget Cure My Stress?" The Times UK. Accessed March 20, 2025. https://www.thetimes.co.uk/article/pulsetto-vagus-nerve-stimulation-device-vctnvfxwr/.

Zanesco, A. P., B. G. King, K. A. Maclean, e C. D. Saron. 2016. "Attentional and Affective Consequences of Technology-Supported Mindfulness Training: A Randomized, Active Control, Efficacy Trial." Psychophysiology. Accessed March 20, 2025. https://pmc.ncbi.nlm.nih.gov/articles/PMC5127005/.

de Manincor, M., Bensoussan, A., Smith, C., Barr, K., & Schweitzer, I. (2015). Individualized yoga for reducing depression and anxiety, and improving well-being: A randomized controlled trial. Depression and Anxiety, 33(9), 816-828. https://doi.org/10.1002/da.22400:contentReference[oaicite:17]{index=17}.

van der Linden, D., Tops, M., & Bakker, A. B. (2021). "The Neuroscience of the Flow State: Involvement of the Locus Coeruleus Norepinephrine System." Neuroscience & Biobehavioral Reviews.